\documentclass[aps,prl,twocolumn,groupedaddress,showpacs]{revtex4}

\usepackage[pdftex]{graphicx}

\begin{document}

\title{Comment on the paper ``Electromagnetic Wave Dynamics in Matter-Wave Superradiant Scattering'' by L. Deng, M.G. Payne, and E.W. Hagley}

\author{Wolfgang Ketterle}

\affiliation{Department of Physics, MIT-Harvard Center for
Ultracold Atoms, and Research Laboratory of Electronics,
Massachusetts Institute of Technology, Cambridge, Massachusetts,
02139}

\date{\today}


\maketitle

The paper by Deng et al. \cite{Deng2010} presents an analytic theoretical description of matter-wave superradiance \cite{Inouye1999} which claims to go beyond previous theoretical frameworks.   I show here that the theory presented in this paper is not a description of superradiance per se,  but rather an elegant perturbative description of a Raman amplifier far away from the superradiant threshold.  As such, it merely is a limiting case of previously known treatments of superradiance.  Two additional new findings of the paper are incorrect:  (1) The claim that adiabatic elimination of the excited state of the atoms is only possible when the probe pulse propagates slowly.  (2) The prediction that superradiance has a dependence on the sign of the detuning of the pump laser due to a phase-matching condition.

 The theory of Raman (or Rayleigh) amplifiers is well-known.  For the situation of matter-wave superradiance in a Bose-Einstein condensate it was summarized in Ref. \cite{Inouye2000}.  If a medium is illuminated with a pump laser beam then there is Raman gain which is described e.g. by equation (1) of Ref. \cite{Inouye2000}.  Since the Raman resonance is narrow, it is accompanied by a narrow dispersive feature that leads to a slow group velocity of the amplified beam \cite{Inouye2000}.  The main result of Ref. \cite{Deng2010}, equation (7), describes  just this phenomenon in the form of a rigorously derived propagation equation for the amplified probe beam.  The only addition is the inclusion of a weak loss term due to off-resonant Rayleigh scattering of the probe beam (parametrized by $\beta_0$ in eq. (7) of [1]) which is completely negligible in the experimental studies.

As described in Ref. \cite{Inouye2000}, superradiance is a non-linear process where the build up of a matter wave grating enhances the Raman gain beyond the perturbative description used in \cite{Deng2010}.  Positive feedback leads then to a run-away situation:  At the threshold for superradiance, the optical Raman gain diverges, and superradiance starts spontaneously without any probe laser input.  The perturbative treatment of Ref. \cite{Deng2010} (which explicitly assumes a classical seed laser field) does not include such feedback, and can therefore not describe the onset of superradiance.  The  threshold in Ref. \cite{Deng2010} called ``superradiantly generated field gain threshold''  is the point at which the perturbative Raman gain exceeds the (negligible) off-resonant absorption of the probe laser beam. However, it has nothing to do with superradiance, and only depends on density.  In contrast, the superradiant threshold depends on optical density, i.e. also on the length of the sample \cite{Inouye1999,Inouye2000}.

All previous treatments of matter-wave superradiant scattering eliminated the electronically excited state since the pump laser field is strongly detuned, typically GHz.  The authors of Ref. \cite{Deng2010} argue that this approximation can only be made when the transit time of the light pulse through the system is longer than the inverse detuning.  This is incorrect: the adiabatic elimination is always possible when the time scale for the process under study is slower than the inverse detuning. Since superradiance occurs on microsecond time scales, this condition is always fulfilled, no matter how slowly the probe light propagates.  In the same context, the authors claim to be the first to describe the slow propagation of light under Raman amplification and superradiance, but this was both described theoretically and experimentally observed in Refs.\cite{Inouye2000} and \cite{Ketterle2001}.

Finally, the authors derive an equation which predicts a finite phase-mismatch for the Raman amplified probe beam, that depends on the sign of the detuning of the probe laser.  The gain process of the Raman amplifier in the current situation is spontaneous Rayleigh scattering.  This spontaneous process always fulfills momentum and energy conservation through the atomic recoil and the frequency of the scattered light.  Therefore, superradiance is ``auto-phase matching''.  The mistake in Ref. \cite{Deng2010} is the assumption that the momentum of the recoiling atoms in the medium is identical to the photon momentum in vacuum, whereas it was shown experimentally that the recoil momentum is modified by the medium \cite{Campbell2005}.  Including this effect guarantees phase matching independently of the pump laser detuning. A recent preprint \cite{Deng_preprint} by two of the authors of Ref. \cite{Deng2010} already acknowledges this.

 Recent experiments by the paper's authors \cite{Deng_preprint} and other groups \cite{DSK} verified the prediction that superradiant Rayleigh scattering depends on the sign of the pump laser detuning.  However, the explanation for this phenomenon in terms of phase matching does not appear valid for the reasons outlined above.  Future explorations of this effect, both experimental and theoretical, are warranted.

 The author thanks D. Schneble, D. Stamper-Kurn, and P. Meystre for discussions.

\bibliographystyle{apsrev}

\end{document}